\documentclass[12pt]{article}

\usepackage{graphicx}
\usepackage{bm}
\usepackage{amsmath}
\usepackage{amssymb,xfrac}
\usepackage[bbgreekl]{mathbbol}
\usepackage{cite}

\def\be{\begin{equation}}
\def\ee{\end{equation}}
\def\bea{\begin{eqnarray}}
\def\eea{\end{eqnarray}}

\begin{document}

\pagestyle{plain}

\begin{center}
~

\vspace{1cm} {\large \textbf{Phase Transition by 0-Branes of U(1) Lattice Gauge Theory}}

\vspace{1cm}

Amir H. Fatollahi

\vspace{.5cm}

{\it Department of Physics, Alzahra University, \\ P. O. Box 19938, Tehran 91167, Iran}

\vspace{.3cm}

\texttt{fath@alzahra.ac.ir}

\vskip .8 cm
\end{center}

\begin{abstract}
The site reduction of U(1) lattice gauge theory is used to model
the 0-branes in the dual theory. The reduced theory
is the 1D plane-rotator model of the angle-valued coordinates on
discrete world-line. The energy spectrum is obtained exactly via the
transfer-matrix method, with a minimum in the lowest energy as a 
direct consequence of compact nature of coordinates. 
Below the critical coupling $g_c=1.125$ and temperature $T_c=0.335$ the system undergoes a first order phase transition between coexistent phases with lower and higher gauge couplings. The possible relation between the model and the proposed role for magnetic monopoles in confinement mechanism based on dual Meissner effect is pointed.
\end{abstract}

\vspace{1cm}


\hfill \texttt{\footnotesize arXiv:1603.04458}

\newpage
According to string theory the gauge fields and coordinates are interchanged
upon the action of T-duality \cite{tasi}. In particular, upon the compactification the gauge fields arising from open strings
would emerge as the transverse coordinates of Dp-branes in the dual
compactified space, leading to the correspondence \cite{tasi,9510017}
\begin{align}\label{1}
A_i &\longleftrightarrow X_i/l_s^2
\end{align}
with $l_s$ as the string theory length; see \cite{cfdual} for another formulation of correspondence between coordinates and gauge fields.
Dp-branes are proposed to represent
the solutions of the effective field theory possessing
charge and mass proportional to the inverse string coupling $\lambda_s$.
The dynamics of coordinates $X_i$'s in the weak coupling limit is captured by the
dimensional reduction of the ordinary U(1) gauge theory to the world-volume of D-brane \cite{tasi,9510017}. In the case of D0-brane,
all spatial components of the gauge field would appear as the time dependent
coordinates \cite{tasi}, resulting
\begin{align}\label{2}
S_0=\int \! dt~\frac{m_0}{2}\, \dot{x}_i^2,
\end{align}
with $m_0\propto 1/g^2 = 1/\lambda_s$ ($g$ as gauge coupling) \cite{tasi}. 
In the case of $N$ Dp-branes the transverse coordinates would appear
as $N$ dimensional hermitian matrices \cite{9510135}.

It is reasonable to ask about the consequences of the correspondence (\ref{1}) proposed by T-duality at strong coupling regime. In this way the lattice gauge theories are the natural candidates, as they have shown their capacity to capture the essential features expected at strong coupling regime \cite{lattice}. As the underlying fundamental string theory of lattice gauge theories is unknown, one may use (\ref{1}) simply in a formal 
way. It is remarkable to note that in the lattice formulation of gauge theories the gauge fields appear to be periodic variables \cite{lattice}, just like the coordinates of Dp-branes in the dual compact theory \cite{tasi}. It is known that treating the gauge fields as compact angle variables, such as those on lattice, reveal very non-trivial aspects of gauge theories \cite{lattice,polya1,polya2,thooft1,kogut}. Accordingly, as a natural expectation by (\ref{1}), it seems reasonable to expect non-trivial aspects when the coordinates appear as angle variables too. In fact, comparing with rather trivial form of (\ref{2}), the lattice action with the angle variable coordinates inserted would seem quite different. Later it will be noticed that in a path-integral representation for compact coordinates, in contrast to infinite extent coordinates, the normalization factor can not be absorbed by a change of integration variable. As a consequence, this would cause that the lowest energy develop a minimum. 

The pure gauge sector of U(1) gauge theory on Euclidean lattice is given by \cite{lattice}:
\begin{align}\label{3}
S_\mathrm{gauge}=\frac{1}{2g^2}\sum_{\vec{n}}\sum_{\mu\nu}
\left( e^{\mathrm{i} f_{\vec{n},\mu\nu}}-1\right)
\end{align}
in which the basic object for each lattice plaquette of size ``$\,a\,$" is defined by
\begin{align}\label{4}
e^{\mathrm{i} f_{\vec{n},\mu\nu}}:=
e^{\mathrm{i}\, aA_{\vec{n},\mu}} e^{\mathrm{i}\,aA_{\vec{n}+\hat{\mu},\nu}}
e^{-\mathrm{i}\, aA_{\vec{n}+\hat{\nu},\mu}} e^{-\mathrm{i}\, aA_{\vec{n},\nu}}.
\end{align}
with $A_{\vec{n},\mu}$ as the gauge field at lattice site $\vec{n}$ in direction $\mu$, and $\hat{\mu}$ as the unit-vector along direction $\mu$. It is assumed $-\pi\leq a\,A \leq\pi$ \cite{lattice}. In the continuum limit $aA\ll 1$, defining  $F_{\vec{n},\mu\nu}:=f_{\vec{n},\mu\nu}/a^2$, the action (\ref{3}) reduces to \cite{lattice}
\begin{align}\label{5}
S_\mathrm{gauge}\simeq  -\frac{1}{4g^2} a^4 \sum_{\vec{n}}F_{\vec{n},\mu\nu}^2
\to  -\frac{1}{4g^2} \int d^4 x\,F_{\mu\nu}^2.
\end{align}
As mentioned earlier, in the formal use of (\ref{1}), after removing the dependence on spatial directions, the components of the gauge fields are interpreted as coordinates. Assuming the following between dimensionless quantities:
\begin{align}\label{6}
a\, A^i\to  x^i/R
\end{align}
leads to
\begin{align}\label{7}
f_{\vec{n},0i}\to (x^i_{n+1}-x^i_{n})/R,~~~~~
\exp({\mathrm{i}\,f_{\vec{n},ij}})\to 1
\end{align}
In above ``$\,n\,$" represents the dependence on the discrete imaginary time, as the only remaining
coordinate of the original space-time lattice. By these, the action (\ref{3}) is reduced to the form
\begin{align}\label{8}
S_{0}=\frac{1}{g^2}
 \sum_{n,i}\left(\cos\frac{x^i_{n+1}-x^i_{n}}{R}-1\right)
\end{align}
which is the sum of copies of the 1D plane-rotator model of magnetic systems.
In fact the close relation between lattice gauge theories and
spin systems was recognized from the first appearance of these theories \cite{lattice,kogut}, and has been used widely for
better understanding the gauge theory side. In particular, the
so-called Villain model \cite{villain}, as an approximation to the plane-rotator model,
was used for gauge theory purposes \cite{banks,savit,guth,spencer,jaffe}.
Here the model is interpreted as a discrete world-line endowed by the
compact coordinates $x^i$'s with
\begin{align}\label{9}
-\pi R &\leq  x^i \leq \pi R
\end{align}
The action (\ref{8}), in contrast to the rather trivial form of (\ref{2})
by ordinary gauge theory, treats the coordinates as angle compact variables. In the first place let us check the continuum limit defined by:
\begin{align}\label{10}
\begin{split}
aA^i&=x^i/R\ll 1\cr
x_{n+1}-x_n&\to a\, \dot{x}\cr
\sum_n&\to a^{-1}\int\! dt
\end{split}
\end{align}
leading to
\begin{align}\label{11}
S_0\simeq \frac{-a}{2g^2R^2}\int dt ~\dot{x}^{2}_i
\end{align}
The above describes the dynamics of a free particle with mass
$m_0=a/(g^2R^2)$ in the imaginary time formalism.
It is mentioned, as far as the dependence on
coupling constant is concerned, the mass corresponds to that
of a 0-brane. Following \cite{lattice} it is useful to define
the new variables
\begin{align}\label{12}
y^i= x^i/R
\end{align}
taking values in $[-\pi,\pi]$. Then setting $\kappa=1/g^2$ 
the action (\ref{8}) takes the form
\begin{align}\label{13}
S_0=\kappa \sum_{n,i}\left(\cos(y^i_{n+1}-y^i_{n})-1\right)
\end{align}
As the action is fully separable for each direction, it is sufficient to consider
only one copy, dropping the index $i$ hereafter.
Following the original prescription introduced for
lattice gauge theories (Sec.~IIIB of \cite{lattice}), the action with discrete
imaginary time can be used to define the quantum theory based on
the transfer-matrix $\hat{V}$, defined by its matrix elements 
between two adjacent times $n$ and $n+1$ \cite{lattice}. Specially for dynamics of a particle it takes the form \cite{wipf}:
\begin{align}\label{14}
\langle  y_{n+1} |\hat{V} |  y_n\rangle= \sqrt{\frac{\kappa}{2\pi}}
\,\exp\left[\kappa \left(\cos( y_{n+1}- y_n)-1\right)\right]
\end{align}
in which the normalization prefactor has to be inserted to match the propagator $\langle x_2,t_2|x_1 ,t_1\rangle \propto \sqrt{m_0}\exp
\left(\frac{-m_0(x_2-x_1)^2}{(t_2-t_1)}\right)$ in the continuum limit \cite{wipf}.  Then the Hamiltonian of the system is related to the transfer-matrix
$\hat{V}$ by \cite{lattice,wipf}
\begin{align}\label{15}
\hat{V}=e^{-a\,\hat{H}}
\end{align}
by which the eigenstates of $\hat{H}$ are those of $\hat{V}$,
with eigenvalues given by \cite{lattice,wipf}
\begin{align}\label{16}
E=-a^{-1} \ln\lambda
\end{align}
where $\lambda$ is the corresponding eigenvalue of $\hat{V}$.
Provided that $\hat{V}$ does not have negative eigenvalues, the
above would give a consistent description of the quantum
theory based on an action with discrete imaginary time \cite{lattice,wipf}.
Here we use the same prescription for 0-branes emerged from lattice gauge theory
as well. First,  using the identity for
the modified Bessel function of the first kind \cite{mattis}:
\begin{align}\label{17}
\exp[\kappa\cos( y'- y)]=\sum_{s=-\infty}^\infty I_s(\kappa) \, e^{\mathrm{i}\,
s\,( y'- y)}
\end{align}
we have for (\ref{14})
\begin{align}\label{18}
\langle  y_{n+1} |\hat{V} |  y_n\rangle= \sum_{s=-\infty}^\infty
 \sqrt{\frac{\kappa}{2\pi}}e^{-\kappa}I_s(\kappa)\, e^{\mathrm{i}\,s\,
( y_n- y_{n+1})}
\end{align}
by which one reads the normalized plane-wave 
$\psi_s(x)=\frac{1}{\sqrt{2\pi}}\exp(\mathrm{i}\,s\, y)$ 
as eigenfunction (recall $-\pi\leq y \leq \pi$) with the eigenvalue
\begin{align}\label{19}
\lambda_s(\kappa)=\sqrt{2\pi\kappa} \, e^{-\kappa}I_s(\kappa) 
\end{align}
By the known properties of $I_s$-functions we have
$\lambda_s=\sqrt{2\pi \kappa}\,e^{-\kappa}I_s(\kappa)\geq 0$.
This guaranties that the transfer-matrix approach defined by (\ref{14})-(\ref{16}) would lead to a consistent quantum 
theory by action (\ref{13}).
Also by $I_s(z)=I_{-s}(z)$ the spectrum is doubly
degenerate for $s\neq 0$. The energy eigenvalues are found by (\ref{16})
and (\ref{19}) 
\begin{align}\label{20}
E_s(\kappa)=-\frac{1}{a} \ln \left[\sqrt{2\pi\kappa} \, e^{-\kappa}I_s(\kappa)  \right]
\end{align}
The behavior of above at zero coupling limit $\kappa=g^{-2} \to\infty$
can be checked by the saddle point approximation for Bessel functions
\begin{align}\label{21}
I_s(\kappa)=\lim_{\kappa\to\infty}\frac{1}{2\pi}\int_{-\pi}^\pi d y~ \exp(\kappa\cos y +\mathrm{i}\,s\, y)
\simeq \frac{e^\kappa}{\sqrt{2\pi \kappa}} \exp\left(-\frac{s^2}{2\kappa}\right)
\end{align}
by which for (\ref{20}) we obtain
\begin{align}\label{22}
E_s\simeq  \frac{s^2}{2a\kappa}
\end{align}
matching the energy $E=p^2/(2m_0)$ of a free particle with 
momentum  $p=s/R$ along the compact direction,
and mass  $m_0=\kappa\, a/R^2$ by (\ref{11}). 
So in the limit
$\kappa=g^{-2} \to\infty$ the spectrum approaches to that of an ordinary particle. For the intermediate coupling the spectrum is discrete.
In the strong coupling limit $\kappa=g^{-2}\to 0$, using
\begin{align}\label{23}
I_s(z)\simeq \frac{1}{s!}\left(\frac{z}{2}\right)^s,~~~~z\ll 1
\end{align}
we have
\begin{align}\label{24}
E_s=(s+\frac{1}{2})\,\frac{\ln g^2}{a}+O(s\ln s) +O(g^{-2})
\end{align}
in which the 2nd term is independent of the coupling constant and
is relevant only for $s\gtrsim \ln g^2\gg 1$. Also at strong coupling
\begin{align}\label{25}
E_{s+1}-E_s\simeq \frac{\ln g^2}{a} \gg \frac{1}{a}
\end{align}
\begin{figure}[t]
	\begin{center}
		\includegraphics[scale=1]{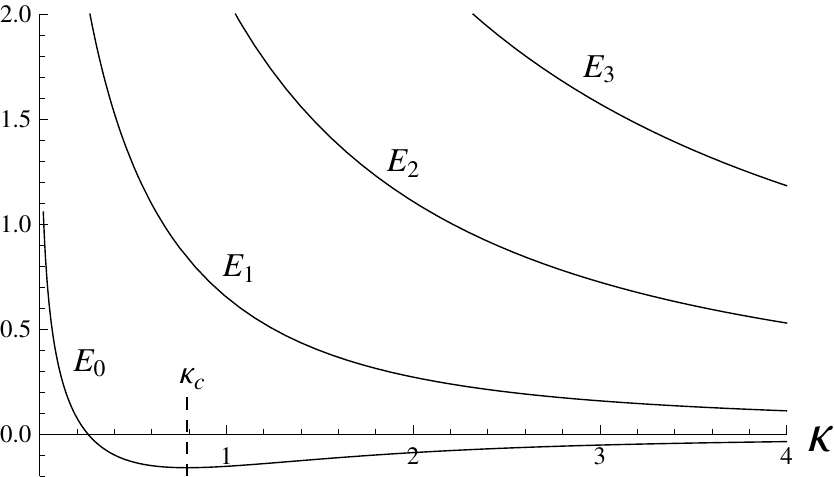}
	\end{center}
	\caption{The few lowest energies by (\ref{20}) versus $\kappa$
		($E$ unit: $a^{-1}$).}
\end{figure}
The interesting observation by the spectrum (\ref{20}) is 
about the energy of ground-state, which has a minimum at $\kappa_c=0.790$, corresponding to coupling $g_c=1/\sqrt{\kappa_c}=1.125$; see Fig.~1. As expected the existence of minimum leads to a 
first order phase transition. The one-particle partition function may be evaluated by the definition
\begin{align}\label{26}
Z_1(\beta,\kappa):=\sum_{s=-\infty}^\infty e^{-\beta\, E_s(\kappa)}
\end{align}
or by means of the transfer-matrix method ($\beta$ in $a$ units) \cite{wipf}
\begin{align}\label{27}
Z_1(\beta,\kappa)=\mathrm{Tr} \,\hat{V}^\beta=
\int_{-\pi}^\pi \prod_{i=1}^{\beta -1}\sqrt{\frac{\kappa}{2\pi}}\, d y_i 
~\exp\left[\kappa\sum_{n=0}^{\beta-1} \left(\cos( y_{n+1}- y_n)-1\right)\right]
\end{align}
supplemented by the periodic condition $y_0=y_\beta$.
In the present case the equivalence of (\ref{26}) and (\ref{27}) is 
checked by numerical evaluations. The basic observation by the compact angle variable in above is, in contrast to the situation with infinite extent coordinates, the normalization factor can not be absorbed by a change of integration variable. As the minimum of $E_0$ is in variable $\kappa$, we need ${M}$ as the thermodynamical conjugate variable, defined by ($T=\beta^{-1}$)
\begin{align}\label{28}
{M}(\beta,\kappa):= T \,\frac{\partial \ln Z_1(\beta,\kappa)}{\partial\; \kappa}
\end{align}
which is also interpreted as the equation-of-state of the system.
The Gibbs free energy can represent the exact nature of the 
phase transition, defined by
\begin{align}\label{29}
G_1=A_1+\kappa\,M
\end{align}
in which $A_1=-T\ln Z_1$ is the free energy per particle. 
\begin{figure}[t]
	\begin{center}
		\includegraphics[scale=1]{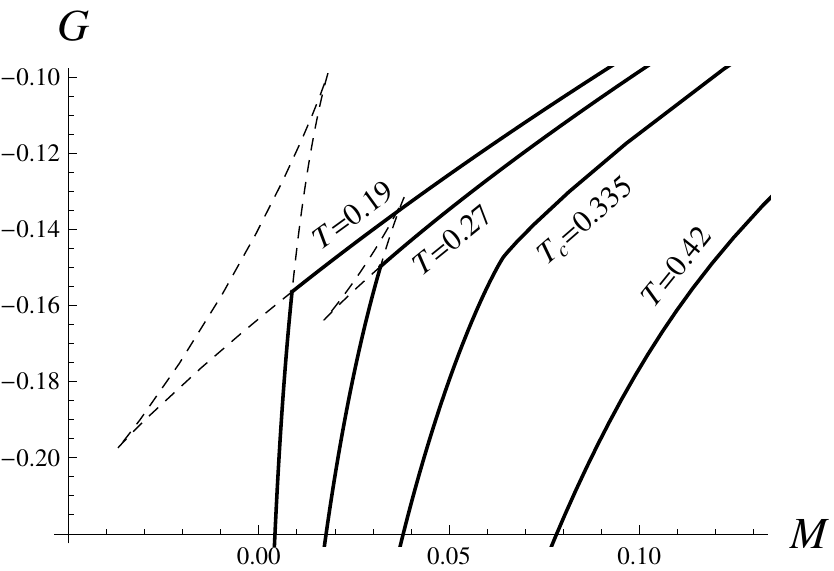}
	\end{center}
	\caption{The $G$-$M$ plots at four temperatures. The dashed pieces are
		not followed by the system due to the minimization of $G$.}
\end{figure}
The isothermal $G$-$M$ plots are presented in Fig.~2. As seen, below the critical temperature $T_c=0.335~a^{-1}$ the plots develop cusps, at which the system follows the path with lower $G$ (solid-lines in Fig.~2), by the minimization of $G$ at equilibrium. As the consequence, for $T<T_c$ there is a jump in first derivative of $\partial G/\partial M$, indicating that the phase transition is a first order one. 
It is apparent by now that the above phase structure is quite similar to 
the gas/liquid transition, for which $G$-$P$ plots show exactly 
the same behavior. In the similar way the equation-of-state (\ref{28}) should be modified 
by the so-called Maxwell construction for $P$-$V$ diagram, 
by which during isothermal condensation the pressure (here $M$) is fixed. 
The results of the Maxwell construction for the present model are plotted as
isothermal $M$-$\kappa$ curves in Fig.~3. The flat part at $T_c$ corresponds to 
values:
\begin{align}\label{30}
T_c=0.335:~~\kappa^*=1.403,~~~M^*=0.064
\end{align}
corresponding to the coupling $g^*=1/\sqrt{\kappa^*}=0.844$. 
\begin{figure}[t]
	\begin{center}
		\includegraphics[scale=1.1]{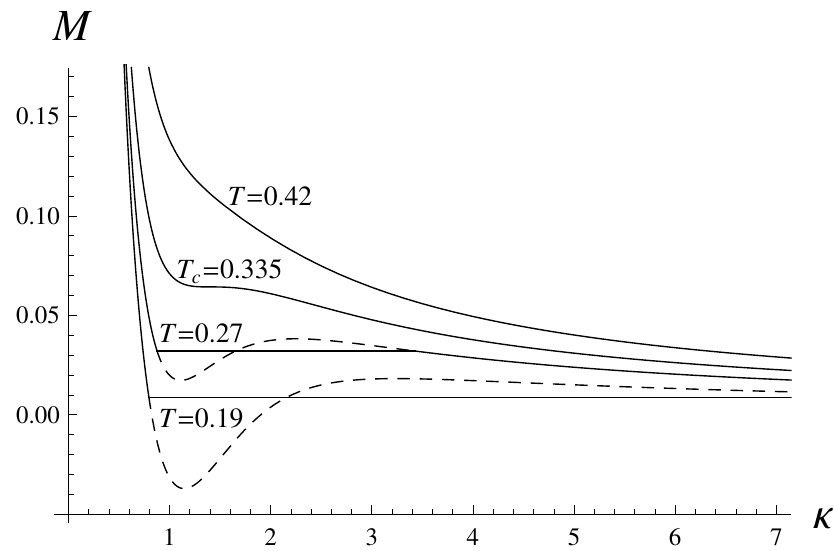}
	\end{center}
	\caption{The isothermal $M$-$\kappa$ plots. The straight-lines are due to the Maxwell construction,
		replacing the dashed parts.}
\end{figure}
For isothermal curves below $T_c$, the straight horizontal parts describe the coexistence phases of lower and higher couplings during the phase transition. 
The interesting fact about the equation-of-states modified by Maxwell construction is that $M$ always remains non-negative, that is $M\geq 0$.  This is specially important by 
expectations from the variable $M$ at weak coupling limit $\kappa\gg 1$, at which the 0-branes behave like ordinary particles. At this limit, back to
(\ref{27}) and (\ref{28}), we have
\begin{align}\label{31}
M \simeq \frac{1}{2} \langle \dot{y}^2 \rangle \propto \frac{T}{m_0}
\end{align}
where the proportionality is by the properties of free ordinary particles. 
In fact the asymptotic tails in Fig.~3 for large $m_0\propto \kappa$ 
are explained by (\ref{31}). 
The behavior (\ref{31}) for ordinary particles is valid for all masses, specially in the small mass limit, leading to the vertical asymptote near $m_0\approx 0$. The 0-branes by the present model also have asymptotes at $\kappa\to 0$, although with a different slope. In fact the main difference between the case with 0-branes in here and that of ordinary particles is about the existence of a phase transition. In particular, by the present model and below the critical temperature $T_c$, the two asymptotes by high and small masses (high and small $\kappa$'s)
are connected with a first order phase transition. 

It would be interesting to see whether the 
present model provides a better understanding of the dynamics of U(1) magnetic monopoles, specially regarding their role in confinement mechanism. Different studies, including those based on lattice formulation of gauge theories, strongly suggest that the Abelain 
U(1) gauge theory has two different regimes, separated by a phase transition. The two phases are supposed to be the confined and Coulomb phases at strong and weak coupling limits, respectively. Both theoretical studies on U(1) lattice gauge theories \cite{lattice,kogut,polya1,polya2,banks,savit,guth,spencer,jaffe} as well as several lattice simulations \cite{creutz,nauen,bha,moria,degran,wiese} have found strong
evidence for such a phase transition. According to the 
mechanism based on a dual version of Meissner effect in superconductors, 
the monopoles have a very distinguished role in such a phase transition \cite{nambu,mandelstam,thooft2}.
Based on the proposed mechanism, at large coupling limit, at which the monopoles have tiny masses, the collective motion of monopoles around the electric fluxes prevents them to
spread, leading to the confinement of the electric charges. 
Instead at small coupling limit, where the monopoles are highly massive, 
the electric fluxes originated from source charges are likely to spread over space, leading to the Coulomb's law. 
It is expected that there is a critical coupling $g_c$ at which the
transition from confined phase to the Coulomb phase occurs.
The lattice simulations suggest $g_c\simeq 1$ \cite{creutz,nauen,bha,moria,degran,wiese}.

It is far from the position to conclude that the present model by the reduced lattice action can give a full explanation for the role of monopoles in the proposed mechanism for confinement. However, one may try to gather pieces of evidences in favor of such explanation.
First, we mention that the effective mass by the model has the same dependence on gauge coupling which is expected for monopoles: $m_0\propto 1/g^2$.
We further mention that the model suggests that the two regimes with low and high coupling constants are related by a first order phase transition. 
The behavior of system at low temperatures, where the main 
contribution to the partition function is by the ground-state, is of particular interest. In the limit $T\to 0$, due to the Maxwell construction, we have $M=0$ for $g<g_c=1.125$. 
So as the consequence of discontinuous nature of first order transitions, at low temperatures and below $g_c$, $\langle v^2\rangle$ is effectively zero; see Fig.~4. This behavior should be compared with (\ref{31}), by which $M$ increases gradually by lowering the mass at constant $T$.
Hence, by the role proposed for the collective motion of monopoles, 
at very low temperatures and below $g_c$ 
the Coulomb phase stay unrivaled with $\langle v^2 \rangle =0$.
On the other hand, exhibiting a high-slope increase of $\langle v^2 \rangle$ at $g_c$, the confined phase at low temperatures should correspond to 
$g > g_c$. This picture and specially the value of critical coupling constant are in agreement with theoretical and numerical studies mentioned earlier. 

\begin{figure}[t]
	\begin{center}
		\includegraphics[scale=1.1]{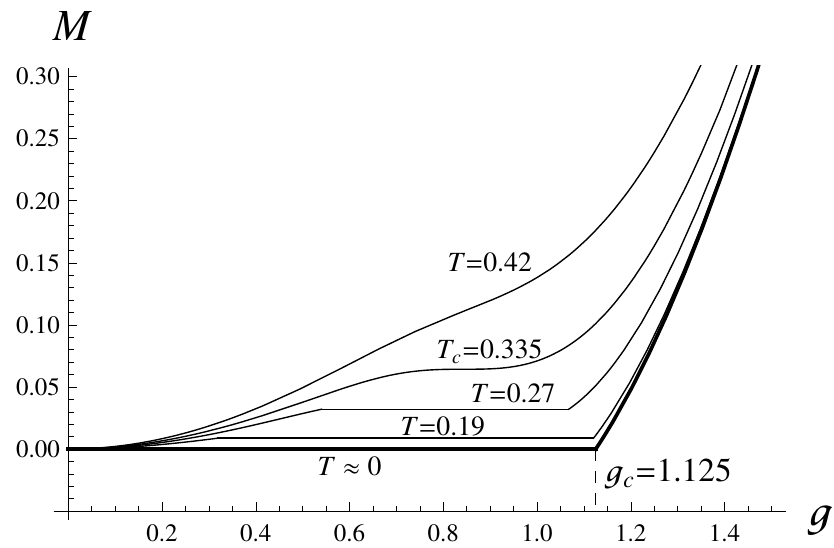}
	\end{center}
	\caption{The isothermal $M$-$g$ plots.}
\end{figure}

As the final point, it is emphasized that the lattice gauge theory and the present model are not belonging to a single theory, though are dually related. In particular, 
in lattice gauge theory the spatial directions are discrete, while gauge fields and 
momenta are compact periodic variables. Instead, the present model is describing a kind of particle dynamics on a space with continuous compact spatial directions, where field fluxes as well as momenta have to be discrete due to compactness of space. These all are rather expected as two theories are related by the correspondence (\ref{1}) suggested by T-duality of string theory. We recall the crucial role of compact angle variable nature 
of dynamical variables in both theories, as a further indication that two sides share common features.

\vspace{2mm}
\textbf{Acknowledgement}: The author is grateful to M.~Khorrami
for helpful discussions on the significance of a minimum in the ground-state.
This work is supported by the Research Council of the Alzahra University.


\end{document}